\documentclass[twocolumn,english,aps,prl,superscriptaddress]{revtex4-1}
\usepackage[T1]{fontenc}
\usepackage[latin9]{inputenc}
\usepackage{prettyref}
\usepackage{amsmath}
\usepackage{amssymb}
\usepackage{graphicx}
\usepackage{esint}

\makeatletter
 \@ifundefined{textcolor}{}
 {%
   \definecolor{BLACK}{gray}{0}
   \definecolor{WHITE}{gray}{1}
   \definecolor{RED}{rgb}{1,0,0}
   \definecolor{GREEN}{rgb}{0,1,0}
   \definecolor{BLUE}{rgb}{0,0,1}
   \definecolor{CYAN}{cmyk}{1,0,0,0}
   \definecolor{MAGENTA}{cmyk}{0,1,0,0}
   \definecolor{YELLOW}{cmyk}{0,0,1,0}
 }

\@ifundefined{definecolor}{\usepackage{color}}{}
\usepackage{babel}

\usepackage{babel}

\usepackage{babel}

\usepackage{babel}

\usepackage{babel}

\makeatother

\usepackage{babel}
\begin{document}

\preprint{This line only printed with preprint option}

\title{Numerically Exact Long Time Magnetization Dynamics at the Nonequilibrium
Kondo Crossover of the Anderson Impurity Model}

\author{Guy Cohen}

\affiliation{Department of Chemistry, Columbia University, New York, New York
10027, U.S.A.}

\affiliation{Department of Physics, Columbia University, New York, New York 10027,
U.S.A.}

\author{Emanuel Gull}

\affiliation{Max Planck Institut für Physik komplexer Systeme, Dresden, Germany}

\affiliation{Department of Physics, University of Michigan, Ann Arbor, MI 48109,
U.S.A.}

\author{David R. Reichman}

\affiliation{Department of Chemistry, Columbia University, New York, New York
10027, U.S.A.}

\author{Andrew J. Millis}

\affiliation{Department of Physics, Columbia University, New York, New York 10027,
U.S.A.}

\author{Eran Rabani}

\affiliation{School of Chemistry, The Sackler Faculty of Exact Sciences, Tel Aviv
University, Tel Aviv 69978, Israel}
\begin{abstract}
We investigate the dynamical and steady-state spin response of the
nonequilibrium Anderson model to magnetic fields, bias voltage, and
temperature using a numerically exact method combining a bold-line
quantum Monte Carlo technique with the memory function formalism.
We obtain converged results in a range of previously inaccessible
regimes, in particular the crossover to the Kondo domain. We provide
detailed predictions for novel nonequilibrium phenomena, including
non-monotonic temperature dependence of observables at high bias voltage
and oscillatory quench dynamics at high magnetic fields. 
\end{abstract}
\maketitle
Strongly correlated open quantum systems appear in a wide variety
of physical situations, including quantum dots in semiconductor heterostructures
\cite{goldhaber-gordon_kondo_1998,Hanson07}, molecular electronics
\cite{park_coulomb_2002,heath_molecular_2003} and the dynamics of
cold atoms \cite{brantut_conduction_2012}. These systems consist
of a finite, interacting region coupled to a continuous set of non-interacting
``bath'' or ``lead'' states which may be maintained at differing
thermodynamic states in such a way that no equilibrium state can exist.
It is natural to describe open systems in terms of quantum impurity
models, which have been used in the description of magnetic impurities
in metals \cite{Anderson61}, the adsorption of atoms on a surface
\cite{Brako81} and as auxiliary problems in the dynamical mean field
approximation to extended lattice systems \cite{Georges96}. More
recently, they have also been of interest in the nonequilibrium context
of mesoscopic transport \cite{meir_low-temperature_1993,rosch_kondo_2001}
and nano-systems coupled to broad leads \cite{Hanson07}.

While attempts are being made to connect nonequilibrium physics to
equilibrium concepts \cite{dutt_effective_2011}, the nonequilibrium
steady state properties of correlated quantum systems continue to
present a formidable challenge to our theoretical understanding. The
main difficulty is that a rigorous evaluation of the long-time and
steady state properties requires an accurate time propagation, starting
from some known initial state and reaching all the way to the steady
state. When this relaxation occurs quickly, a range of powerful semi-analytical
\cite{Langreth91,Fujii03,Eckstein09} and numerical methods \cite{Muhlbacher08,Bulla08,Anders08,Werner09_noneq,Schiro09,HeidrichMeisner09,gull_numerically_2011,cohen_memory_2011-1}
are applicable. However, dynamics in strongly correlated systems may
exhibit a separation of timescales---for example, the spin-relaxation
dynamics in the Kondo regime of a quantum dot are orders of magnitude
slower than those of the corresponding charge relaxation. Existing
theoretical methods are unable to resolve these timescales reliably.

In this Letter we show that a combination of bold-line diagrammatic
Monte Carlo methods \cite{gull_bold-line_2010,gull_numerically_2011}
and the memory-function approach \cite{cohen_memory_2011-1} enables
us to significantly extend the time regime accessible and can, in
some cases, access steady state information within the Kondo regime.
The method is numerically exact and provides unbiased error estimates.
While the calculations presented here for the single impurity Anderson
model, a minimal model for strong interactions in the presence of
baths, the methodology is applicable to any quantum impurity model.

The Anderson impurity model is defined by the Hamiltonian 
\begin{equation}
H=H_{S}+H_{B}+V,\label{eq:hamiltonian}
\end{equation}
 where $H_{S}$ describes the interacting system (or dot) part, $H_{B}$
the non-interacting bath (or leads) part, and $V$ the system--bath
coupling part: 
\begin{eqnarray}
H_{S} & = & \sum_{i=\uparrow\downarrow}\varepsilon_{i}d_{i}^{\dagger}d_{i}+Ud_{\uparrow}^{\dagger}d_{\uparrow}d_{\downarrow}^{\dagger}d_{\downarrow},\\
H_{B} & = & \sum_{k,i=\uparrow\downarrow}\varepsilon_{ik}a_{ik}^{\dagger}a_{ik},\\
V & = & \sum_{k,i=\uparrow\downarrow}t_{ik}d_{i}a_{ik}^{\dagger}+t_{ik}^{*}a_{ik}d_{i}^{\dagger}.\label{eq:coupling-hamiltonian}
\end{eqnarray}
 Here $\uparrow$ and $\downarrow$ represent electronic spin, the
$d_{i}$ and $d_{i}^{\dagger}$ are fermionic system operators for
dot states with energy $\varepsilon_{i}$, $a_{ik}$ and $a_{ik}^{\dagger}$
are fermionic lead operators with energy $\varepsilon_{ik}$ and the
$t_{ik}$ are coupling constants. $k$ is an index iterating over
the lead states. Both the $\varepsilon_{ik}$ and the $t_{ik}$ are
fully defined by the system--lead coupling density $\Gamma\left(\varepsilon\right)\equiv2\pi\sum_{k}\left|t_{k}\right|^{2}\delta\left(\varepsilon-\varepsilon_{ik}\right)$.

Refs.~\cite{nakajima_quantum_1958,zwanzig_ensemble_1960,mori_transport_1965}
have shown that the reduced density matrix $\sigma\left(t\right)=\mathrm{Tr}_{B}\left\{ \rho\left(t\right)\right\} $
($\rho\left(t\right)$ being the full density matrix and $\mathrm{Tr}_{B}\left\{ ...\right\} $
denoting a trace over all bath degrees of freedom) of any system of
the form of Eq.~\prettyref{eq:hamiltonian} exactly obeys the Nakajima--Zwanzig--Mori
equation 
\begin{equation}
i\hbar\frac{\mathrm{d}\sigma\left(t\right)}{\mathrm{d}t}=\mathcal{L}_{H_{S}}\sigma\left(t\right)+\vartheta\left(t\right)-\frac{i}{\hbar}\int_{0}^{t}\mathrm{d}\tau\,\kappa\left(\tau\right)\sigma\left(t-\tau\right).\label{eq:sigma_EOM}
\end{equation}
 Here, the Liouvillian superoperator $\mathcal{L}_{H_{S}}A\equiv\left[H_{S},A\right]$
denotes a commutation with the system Hamiltonian $H_{S}$, with the
same notation defining $\mathcal{L}_{V}$ and $\mathcal{L}_{H}$;
$\vartheta\left(t\right)$ is an initial correlation term which vanishes
for factorized initial conditions $\rho\left(0\right)\equiv\rho_{B}\otimes\sigma\left(0\right)$;
and $\rho_{B}$ is the initial bath density matrix. $\kappa$ is known
as the memory kernel and may be obtained by solving \cite{zhang_nonequilibrium_2006}
\begin{eqnarray}
\kappa\left(t\right) & = & i\hbar\dot{\Phi}\left(t\right)-\Phi\left(t\right)\mathcal{L}_{S}+\frac{i}{\hbar}\int_{0}^{t}\mathrm{d}\tau\Phi\left(t-\tau\right)\kappa\left(\tau\right),\label{eq:kappa_EOM}
\end{eqnarray}
 where the superoperator $\Phi\left(t\right)\equiv\mathrm{Tr}_{B}\left\{ \mathcal{L}_{V}e^{-\frac{i}{\hbar}\mathcal{L}_{H}t}\rho_{B}\right\} $
must in general be obtained from a many body computation whose expense
rapidly increases as $t$ increases. Evaluation of $\Phi\left(t\right)$
for $t$ up to a cutoff time $t_{c}$ allows an exact evaluation of
$\kappa\left(t<t_{c}\right)$. Setting $\kappa\left(t>t_{c}\right)=0$
defines the cutoff approximation, whose convergence may be monitored
from the dependence of results on $t_{c}$ as $t_{c}$ is increased.
In the case of the Anderson impurity model, Ref.~\cite{cohen_memory_2011-1}
has shown that if one is only interested in evaluating the diagonal
elements of the density matrix, all the supermatrix elements $\Phi_{ij,qq^{\prime}}\equiv\left(\left|i\right\rangle \left\langle j\right|\right)^{\dagger}\Phi\left|q\right\rangle \left\langle q^{\prime}\right|$
of $\Phi$ having $i\neq j$ or $q\neq q^{\prime}$ can be set to
zero, with the remaining elements determined by: 
\begin{eqnarray}
\Phi_{ii,qq} & = & \delta_{i0}\left(\varphi_{qq}^{\left(1\right)}+\varphi_{qq}^{\left(3\right)}\right)+\delta_{i1}\left(\varphi_{qq}^{\left(2\right)}-\varphi_{qq}^{\left(3\right)}\right)\nonumber \\
 & + & \delta_{i2}\left(-\varphi_{qq}^{\left(1\right)}+\varphi_{qq}^{\left(4\right)}\right)+\delta_{i3}\left(-\varphi_{qq}^{\left(2\right)}-\varphi_{qq}^{\left(4\right)}\right),\label{eq:Phi_anderson}\\
\varphi_{qq}^{\left(m\right)}\left(t\right) & = & 2i\Im\sum_{k}\mathrm{Tr}_{B}\left\{ \rho_{B}\left\langle q\right|A_{k}^{\left(m\right)}\left(t\right)\left|q\right\rangle \right\} ,\label{eq:anderson_phi}
\end{eqnarray}
 where $A_{k}^{\left(1\right)}=t_{\uparrow k}d_{\uparrow}d_{\downarrow}d_{\downarrow}^{\dagger}a_{\uparrow k}^{\dagger}$,
$A_{k}^{\left(2\right)}=t_{\uparrow k}d_{\uparrow}d_{\downarrow}^{\dagger}d_{\downarrow}a_{\uparrow k}^{\dagger}$,
$A_{k}^{\left(3\right)}=t_{\downarrow k}d_{\uparrow}d_{\uparrow}^{\dagger}d_{\downarrow}a_{\downarrow k}^{\dagger}$
and $A_{k}^{\left(4\right)}=t_{\downarrow k}d_{\uparrow}^{\dagger}d_{\uparrow}d_{\downarrow}a_{\downarrow k}^{\dagger}$.

The evaluation of the $\varphi_{qq}^{\left(m\right)}\left(t\right)$
has previously been performed with real time path integral Monte Carlo
(RT-PIMC) methods \cite{Muhlbacher08,werner_diagrammatic_2008,cohen_memory_2011-1},
revealing that, in the presence of strong electronic correlations,
the memory kernel may develop long tails. Near the Kondo regime this
effect becomes particularly pronounced, making it impossible to converge
the cutoff approximation and highlighting the need for methods able
to obtain $\kappa$ for longer times. Here we show that the problem
can to a large extent be solved by using the bold expansion for impurity
models \cite{gull_bold-line_2010}, a technique related to bold-line
methods for lattice systems \cite{prokofev_bold_2007,prokofev_bold_2008,vanhoucke_bold_2012}.
The bold expansion is based on a stochastic Monte Carlo sampling of
diagrammatic corrections to the propagators obtained from an infinite
partial summation, rather than a sampling of all diagrams. The resulting
procedure converges at lower expansion order and greatly reduces the
severity of the dynamic sign problem, in practice more than doubling
the accessible time scales. Even with bold methods, a direct description
of the slow spin dynamics remains out of reach---however, the bold
method does allow converged access to such dynamics within the memory
formalism.

\begin{figure}
\includegraphics[width=8.6cm]{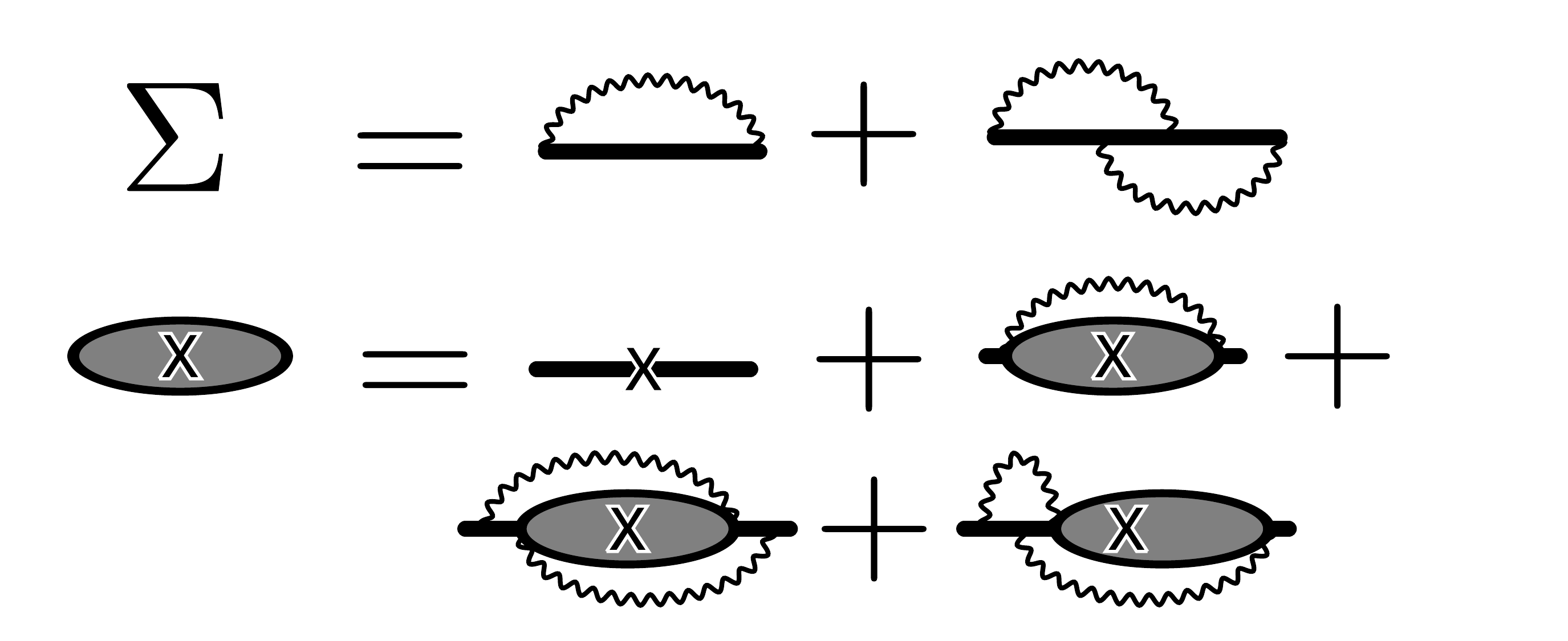}

\caption{Self energies and vertex equations used within the OCA based bold
expansion. Solid lines represent bare propagators, bold lines are
dressed propagators, wavy lines are hybridization interactions and
shaded regions are vertex functions. The vertices are defined on the
unfolded Keldysh contour, such that the final time on the contour
is marked by the central ``X'' and both edges of the contour stand
for the initial time. The NCA is obtained by taking only the first
diagram in the self energy and the first two diagrams in the vertex.\label{fig:diagrams}}
\end{figure}

In the nonequilibrium case, diagrams must be evaluated on the Keldysh
contour. This had previously been done with ``BoldNCA'', using the
non-crossing approximation (NCA) as the underlying partial summation,
but in the course of this work it has become necessary to employ a
``BoldOCA'' built on the more precise one-crossing approximation
(OCA) \cite{gull_bold-line_2010,pruschke_anderson_1989}. Fig.~\ref{fig:diagrams}
illustrates this in diagrammatic terms: the bold-line propagators
and vertex functions (which allow for the summation over hybridization
lines connecting pairs of times on the two different halves of the
Keldysh contour) are obtained from the solution of the NCA or OCA
equations, and used in an expansion which samples diagrams of all
crossing orders. Unbiased error estimates are obtained by jackknife
analysis on data from multiple, uncorrelated Monte Carlo runs (typically
5--10).

We assume a wide, flat band $\Gamma_{i}\left(E\right)=\Gamma_{i}^{L}\left(E\right)+\Gamma_{i}^{R}\left(E\right)$
with $\Gamma_{i}^{L\left(R\right)}\left(E\right)=\frac{\Gamma/2}{\left(1+e^{\nu\left(E-\varepsilon_{c}\right)}\right)\left(1+e^{-\nu\left(E+\varepsilon_{c}\right)}\right)}$;
here $\varepsilon_{c}$ and $\nu$ are the band cutoff energy and
its inverse cutoff width, and $L$ and $R$ are respectively left
and right lead indices. We restrict our calculations to the symmetric
Anderson impurity model in a magnetic field $h$, setting $\varepsilon_{i}=-\frac{U}{2}\pm\frac{h}{2}$
(the formalism is more general and does not rely on this symmetry).
We choose $\Gamma$ as our energy unit, and throughout the rest of
this paper set $U=5\Gamma$, $\varepsilon_{c}=10\Gamma$ and $\Gamma\nu=10$.
The initial conditions are determined by assuming an initially decoupled
system, having left and right leads thermally equilibrated at a temperature
$\beta$ and chemical potentials $\mu_{L}=\frac{V}{2}$ and $\mu_{R}=-\frac{V}{2}$,
respectively. This defines the lesser and greater hybridization functions
$\Delta_{L\left(R\right)}^{<}\left(\omega\right)=-if_{L\left(R\right)}\left(\omega\right)\Gamma_{L\left(R\right)}\left(\omega\right)$
and $\Delta^{>}\left(\omega\right)=i\left(1-f_{L\left(R\right)}\left(\omega\right)\right)\Gamma_{L\left(R\right)}\left(\omega\right)$,
which depend on the temperature and chemical potentials through the
Fermi occupation function $f_{L\left(R\right)}\left(\omega\right)=\frac{1}{1+e^{\beta\left(\omega-\mu_{L\left(R\right)}\right)}}$.
At these parameters, the Kondo temperature is given by $\Gamma\beta_{K}\equiv\frac{\Gamma}{T_{K}}\simeq3.4$
\cite{hewson_kondo_1993}.

\begin{figure}
\includegraphics[width=8.6cm]{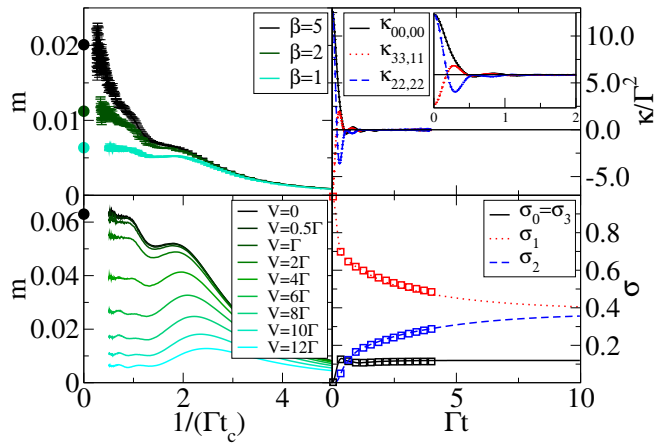}

\caption{Top left panel: The steady-state magnetization obtained from the memory
formalism at several temperatures plotted as a function of the inverse
cutoff time, and compared in the equilibrium cases to exact CT-QMC
results shown as circles, for $h=0.01\Gamma$ and $V=0$. Bottom left
panel: The same plot at $\beta\Gamma=1$ and $h=0.1\Gamma$, for several
voltages. Right panels: equilibrium memory kernel $\kappa$ (top)
and populations $\sigma_{i}$ (bottom) as a function of time for $\beta\Gamma=1$
and $h=0.01\Gamma$. The inset shows the memory kernel at short times.
The squares in the bottom right panel are approximate OCA results.\label{fig:convergence_temperature_and_OCA_comparison}}
\end{figure}

In equilibrium, the magnetization can be evaluated exactly on the
Matsubara axis \cite{werner_continuous-time_2006,gull_bold-line_2010,gull_continuous-time_2011}.
As this magnetization corresponds to the steady state magnetization
at zero bias voltage, it is useful as a benchmark. The top left panel
of Fig.~\ref{fig:convergence_temperature_and_OCA_comparison} displays
the steady state magnetization predicted by the proposed method at
$V=0$, plotted against the inverse cutoff time $\frac{1}{\Gamma t_{c}}$
at several temperatures. Comparing the predicted steady state to the
equilibrium results (circles) shows that the cutoff approximation
converges rapidly even as one crosses the edge of the Kondo regime.
For the very small magnetic field $h=0.01\Gamma$ in Fig.~\ref{fig:convergence_temperature_and_OCA_comparison},
the relative errors are rather large, but considered on the full scale
of the magnetization the precision demonstrated here is remarkable.

The effects of taking the system out of equilibrium are illustrated
in the lower left panel of Fig.~\ref{fig:convergence_temperature_and_OCA_comparison}.
Here a constant temperature $\beta\Gamma=1$ is maintained while the
bias voltage $V$ is varied at $h=0.1\Gamma$; the numerically exact
$V=0$ result is also shown. This plot clearly illustrates that convergence
of the method generally occurs at even shorter times in nonequilibrium
conditions---equilibrium exhibits the longest memory, consistent with
expectations.

An independent approach to verifying convergence relies on direct
examination of individual elements of the memory kernel as a function
of time. Several representative elements are displayed at $h=0.01\Gamma$
and $\beta\Gamma=1$ in the top right panel of Fig.~\ref{fig:convergence_temperature_and_OCA_comparison},
with the inset highlighting short times. Within the times accessible
by BoldOCA, the memory kernel elements go to zero within the numerical
accuracy. Below this, on the same time scale and for the same parameters,
the time dependence of the three distinct elements of the reduced
density matrix $\sigma$ is plotted for an initially magnetized dot
in the lower right panel of Fig.~\ref{fig:convergence_temperature_and_OCA_comparison}.
With this initial condition and within the symmetric Anderson impurity
model, the diagonal density matrix entries $\sigma_{0}$ and $\sigma_{3}$,
which express charging dynamics, are identical. They both relax rapidly,
and in fact their steady state values could have been obtained to
very good accuracy within BoldNCA only. The difference in scale between
the spin relaxation time of $\sigma_{1},\sigma_{2}$ and the memory
decay in the upper panel, however, is striking---and is why our memory
kernel methods are essential for obtaining long-time behavior. To
obtain a reasonable converged steady state directly, one would need
to reach times $\Gamma t\gtrsim20$ with errors of similar magnitude
compared to what we have obtained at $\Gamma t=2$ with the current
approach. The exponential scaling in time typical of all general exact
methods makes this unfeasible.

\begin{figure}
\includegraphics[width=8.6cm]{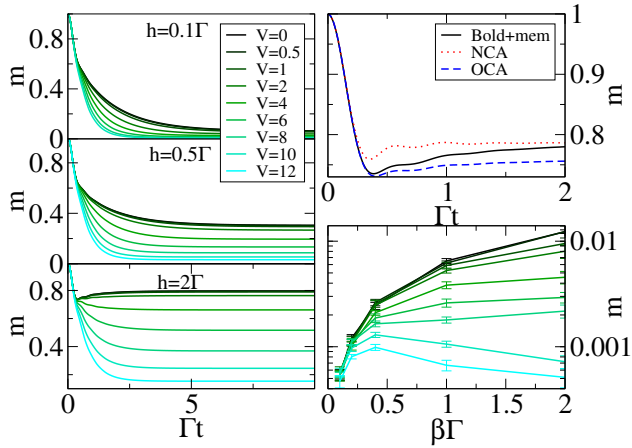}

\caption{Left panels: Time dependence of the cutoff-converged magnetization
at $\beta\Gamma=1$ starting from a fully magnetized dot, at different
magnetic fields $h$ and bias voltages $V$. Top right: Comparison
with NCA and OCA at $h=2\Gamma$ and $V=0$. Bottom right: Temperature
dependence of the $h=0.01\Gamma$ steady state magnetization at different
voltages.\label{fig:voltage_effect_on_convergence_and_memory}}
\end{figure}

We now turn from the demonstration of convergence to presentation
of results. The left panels of Fig.~\ref{fig:voltage_effect_on_convergence_and_memory}
show the time evolution of the magnetization from an initially polarized
state at different voltages and magnetic fields, with $\beta\Gamma=1$.
At low voltages two separate relaxation timescales are apparent: immediate
fast relaxation followed by later slow relaxation. At high enough
fields (bottom), an overshoot effect appears along with oscillatory
behavior which is seen more clearly in the upper right panel. As we
increase the voltage the second timescale is suppressed and eventually
the relaxation becomes exponential. However, the voltage required
in order to reach this regime is surprisingly large. In the top right
panel of Fig.~\ref{fig:voltage_effect_on_convergence_and_memory},
we show that nonequilibrium NCA and OCA (not supplemented by QMC)
do very poorly away from $h=0$ and cannot be considered reliable,
whereas the memory approach and the underlying BoldOCA continue to
converge very well.

In the lower right panel of Fig.~\ref{fig:voltage_effect_on_convergence_and_memory},
we show an example of the temperature dependence of the $t\rightarrow\infty$
limit of the magnetization at constant magnetic field and a range
of bias voltages. Interestingly, at higher voltages (but substantially
below $\frac{V}{2}\approx\omega_{c}$ where the lead chemical potentials
approach the band cutoff) the temperature dependence becomes non-monotonic.
We believe this is a population switching effect \cite{cohen_negative_2008},
which leads to a suppression of the magnetization by population transfer
from the magnetized $\left|1\right\rangle $ and $\left|2\right\rangle $
states to the unmagnetized $\left|0\right\rangle $ and $\left|3\right\rangle $
states which are activated for $V\gtrsim U$. The rate for this transfer
process is approximately proportional to the lead occupation at the
energy difference between the states: $f\left(\beta,\Delta E,\mu\right)=\frac{1}{1+e^{\beta\left(\Delta E-\mu\right)}}$,
with $\Delta E$ equal to half the interaction energy $\frac{U}{2}$
and $\mu=\frac{V}{2}$ or $-\frac{V}{2}$, depending on the lead.
$f$ is therefore an increasing function of temperature for $V<U$
and a decreasing one for $V>U$. At small voltages the effect of the
population transfer results in a reduction of the magnetization (as
expected), while at large voltages the population transfer enhances
the intermediate-temperature magnetization. At still larger temperatures,
the nonequilibrium effects are washed away and normal thermal suppression
of the magnetization occurs. A maximum can therefore be expected,
as is indeed observed in the numerically converged calculations.

\begin{figure}
\includegraphics[width=8.6cm]{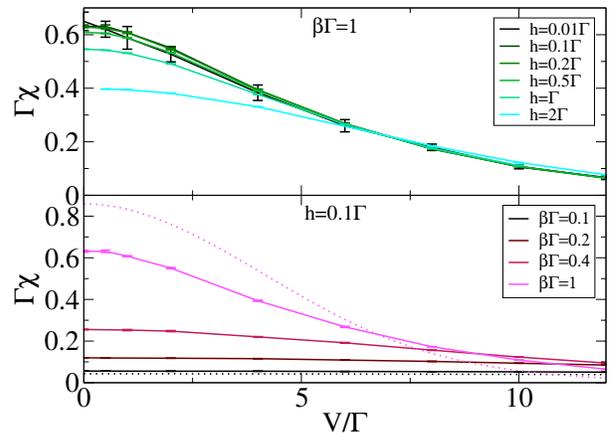}

\caption{Generalized magnetic susceptibility $\chi\equiv\frac{m}{h}$ as a
function of voltage, for (top) different magnetic fields and (bottom)
different temperatures, at $h=0.1\Gamma$. Approximate results from
a master equation calculation are shown in dotted lines for the lowest
and highest temperatures in the lower panel. \label{fig:magnetization_statics_and_dynamics_in_voltage}}
\end{figure}

In Fig.~\ref{fig:magnetization_statics_and_dynamics_in_voltage}
we display the steady state voltage dependence of the generalized
magnetic susceptibility $\chi\equiv\frac{m}{h}$. At small $h$ this
quantity is $h$-independent. The top panel shows clearly how the
regime in which $m$ is linear in $h$ depends on voltage at a constant
temperature. The bottom panel of Fig.~\ref{fig:magnetization_statics_and_dynamics_in_voltage}
shows the voltage dependence at different temperatures within the
linear regime. One immediately noticeable feature is the decrease
of $\chi$ with increasing $\beta$ at high voltage, which corresponds
to the non-monotonic temperature dependence discussed in the bottom
panel of Fig.~\ref{fig:magnetization_statics_and_dynamics_in_voltage}.
A second interesting feature is the fact that the plots have a simple,
Lorentzian-like structure, suggesting that the results may be in a
regime accessible to analytical methods based on performing logarithmic
corrections around rate equations \cite{paaske_nonequilibrium_2004}:
in the dotted lines in the bottom panel we show for comparison results
obtained by solving the classical rate equations (obtained by simple
perturbation theory to second order in the hybridization). The large
discrepancy between the master equation and numerically exact results
at $\beta\Gamma=1$ shows that in the crossover regime, even at large
voltage a correct account of the physics requires accurate calculations
of the sort performed here.

In conclusion, by unifying numerically exact bold Monte Carlo methods
with the exact memory approach we have developed a new, numerically
exact formalism free from systematic errors and well suited for the
real time solution of nonequilibrium quantum impurity models. In practice,
the capabilities of this formalism are unparalleled: the method generates
precise, converged results at all timescales, in cases where the current
state-of-the-art approximate methods clearly fail. For the nonequilibrium
Anderson impurity model, the formalism performs well even as one enters
the Kondo regime, a regime previously inaccessible with accurate numerical
methods.

Our formalism has allowed us to explore the detailed behavior of the
nonequilibrium magnetization, and we have made predictions regarding
multi-scale, oscillatory quenching dynamics at high magnetic fields;
the effect of voltage on dynamical relaxation; and population-driven
reversal of the magnetization's temperature dependence at high voltages.
These results are obtained at parameters where no other currently
available method is reliable. As the temperature is further lowered,
one expects to encounter the formation of Kondo peaks at the chemical
potential. How this will affect the behavior described here remains
an interesting and open question, and work is currently being carried
out to further investigate this issue. Future research will address
lower temperatures and a wider variety of observables; it is also
worth stressing that both bold techniques and the memory formalism
are not specific to the Anderson impurity model, and are expected
to have many more applications. 
\begin{acknowledgments}
GC is grateful to the Azrieli Foundation for the award of an Azrieli
Fellowship and to the Yad Hanadiv--Rothschild Foundation for the award
of a Rothschild Postdoctoral Fellowship and acknowledges the hospitality
of MPI PKS Dresden and Columbia University. This work was supported
by the US--Israel Binational Science Foundation and by the FP7 Marie
Curie IOF project HJSC. DRR acknowledges NSF CHE-1213247. AJM acknowledges
NSF DMR 1006282. Our implementations were based on the ALPS \cite{bauer_alps_2011}
libraries. 
\end{acknowledgments}
 \bibliographystyle{apsrev4-1}
\bibliography{Library}

\end{document}